# AI Guided Early Screening of Cervical Cancer

Dharanidharan S I, Suhitha Renuka S V, Ajishi Singh, Sheena Christabel Pravin*

School of Electronics Engineering,

Vellore Institute of Technology, Chennai.

*sheenachristabel.p@vit.ac.in

## Abstract

In order to support the creation of reliable machine learning models for anomaly detection, this project focuses on preprocessing, enhancing, and organizing a medical imaging dataset. There are two classifications in the dataset: normal and abnormal, along with extra noise fluctuations. In order to improve the photographs' quality, undesirable artifacts, including visible medical equipment at the edges, were eliminated using central cropping. Adjusting the brightness and contrast was one of the additional preprocessing processes. Normalization was then performed to normalize the data. To make classification jobs easier, the dataset was methodically handled by combining several image subsets into two primary categories: normal and pathological. To provide a strong training set that adapts well to real-world situations, sophisticated picture preprocessing techniques were used, such as contrast enhancement and real-time augmentation (including rotations, zooms, and brightness modifications). To guarantee efficient model evaluation, the data was subsequently divided into training and testing subsets. In order to create precise and effective machine learning models for medical anomaly detection, high-quality input data is ensured via this thorough approach. Because of the project pipeline's flexible and scalable design, it can be easily integrated with bigger clinical decision-support systems.

## 1. Introduction

Recent developments in artificial intelligence and machine learning have revolutionized medical imaging, allowing for more precise and effective illness diagnosis. The caliber and structure of the data utilized to train these models provide the basis of these technological developments. Data imbalance, noise, and imaging condition variability are some of the distinctive difficulties that medical imaging datasets, especially those utilized for anomaly identification, frequently face. To overcome these obstacles and guarantee the creation of reliable machine learning models, the data must be carefully preprocessed, augmented, and organized.

Images in medical imaging datasets are usually classified into classes like normal and abnormal, with variances resulting from environmental noise, patient-specific factors, and imaging equipment. Effective preprocessing of these datasets is crucial for guaranteeing the precision and dependability of machine learning models. Noise reduction, cropping to eliminate superfluous regions, normalization to uniformize pixel intensities, and augmentation to artificially boost the training data's diversity are all examples of preprocessing techniques. By taking these actions, the data quality is improved and the models' capacity to generalize to new, untested data is increased.

In order to remove unwanted background details or medical instruments from medical photos, central cropping is one of the crucial preprocessing techniques used in this research. The danger of adding artifacts to the training data is reduced by concentrating on the image's center, which usually contains the area of interest. Normalization standardizes the pixel intensity values to a constant range, which facilitates the interpretation of the data by machine learning models. Other preprocessing procedures, such as brightness and contrast modifications, provide consistency throughout the dataset.

The inherent drawbacks of medical datasets, such small sample numbers and class imbalances are greatly mitigated by data augmentation. Rotations, zooming, flipping, brightness modifications, and other augmentation techniques create a more complete training dataset by simulating a variety of imaging circumstances. This procedure improves the machine learning model's robustness and reduces overfitting, allowing it to function well in a range of real-world scenarios. This project's main goal is to create a high-quality dataset that will act as the foundation for machine learning algorithms that can identify abnormalities in medical images. This effort guarantees the scalability and dependability of these models in clinical contexts by utilizing sophisticated preprocessing and augmentation procedures. The ultimate goal of this project's output is to enhance patient outcomes by enabling the early diagnosis of anomalies and adding to the expanding corpus of work in medical imaging and machine learning.

## 2.Related Work

Gomes et al. in This study emphasizes the revolutionary possibilities of combining machine learning (ML) and deep learning (DL) approaches for the identification and categorization of cervical cancer. With the help of Simple Logistic classifiers and pre-trained models like ResNet152, the study attains an impressive 98.08% accuracy rate. In order to highlight the significance of sophisticated computational tools in boosting medical diagnoses, I can include this in my paper along with the benefits of hybrid DL-ML techniques for increasing diagnostic precision.

Vargas-Cardona et al. in This study explores the application of artificial intelligence (AI) to early cervical cancer detection using imaging. It looks closely at the diagnosis accuracy of a number of AI algorithms that are used to analyze images from digital colposcopy, cervicography, and mobile devices. Among the thirty-two publications published between 2009 and 2022, support vector machines (SVM) and deep learning algorithms (e.g., CNN, ResNet, VGG) were the most utilized AI techniques. With accuracies over 97%, these AI approaches demonstrated exceptional diagnostic performance. Although more research is needed to validate these positive findings, the study highlights the growing use of AI in cervical cancer screening.

This study by Jue Wang et al. (2024) introduces the AI-based cervical cancer screening system (AICCS), which evaluates cervical cytology images to enhance early detection. The method employs AI models to detect abnormal cells at the patch level and categorizes entire slides into various cytology groups. It performed exceptionally well on a range of datasets, including prospective and randomized trials, with an AUC of 0.947, a sensitivity of 94.6%, and a specificity of 89.0%.

Wei Wang et al.'s (2022) study used a targeted literature review (TLR) and a systematic literature review (SLR) to examine cervical cancer screening methods and guidelines in 11 different countries, including the US, Canada, and several European and Asia-Pacific countries. The authors examined peer-reviewed papers and policy documents from January 2005 to January 2021 using databases like Embase, Medline, and Cochrane as well as gray literature from websites of government and health authorities. The study focused on nations having established screening systems, pilot programs, and easily accessible data on screening practices, all of which met the PICOS criteria for relevance. Except for Japan, where only English-language documents were taken into consideration, the materials were evaluated by a multinational team of reviewers. This study aimed to provide a comprehensive understanding of global cervical cancer screening guidelines and practices to enhance prevention strategies.

In order to evaluate the application of artificial intelligence (AI)-based algorithms for diagnosing cervical cancer from visual inspection with acetic acid (VIA) images, Roser Viñals et al. (2023) did a systematic literature review (SLR). The review, which was conducted using databases like PubMed, Google Scholar, and Scopus, concentrated on articles released between January 2015 and July 2022. Using histopathology as the gold standard for diagnosis, the studies that evaluated AI algorithms for differentiating between positive (CIN2+) and negative (normal or benign) cervical lesions were chosen for inclusion. Eleven chosen papers were thoroughly examined by the authors, who looked at variables like algorithm types, acquisition tools, dataset properties, and performance metrics like specificity, sensitivity, and accuracy.

To evaluate the application of artificial intelligence (AI)-based algorithms for diagnosing cervical cancer from visual inspection with acetic acid (VIA) images, Roser Viñals et al. (2023) did a systematic literature review (SLR). The review, which was conducted using databases like PubMed, Google Scholar, and Scopus, concentrated on articles released between January 2015 and July 2022. Using histopathology as the gold standard for diagnosis, the studies that evaluated AI algorithms for differentiating between positive (CIN2+) and negative (normal or benign) cervical lesions were chosen for inclusion. Eleven chosen papers were thoroughly examined by the authors, who looked at variables like algorithm types, acquisition tools, dataset properties, and performance metrics like specificity, sensitivity, and accuracy. Using the QUADAS-2 technique, they also conducted a quality evaluation to determine study applicability and bias risk. A qualitative comparison of the preprocessing methods and dataset characteristics employed in the investigations was presented by the research.

Using pap smear images from the SIPaKMeD dataset, which contains both normal and abnormal cell types, Sandeep Kumar Mathivanan et al. (2024) present an automated method for cervical cancer identification. The dataset undergoes further classification within the normal class after being converted into a two-class system (normal and aberrant). To extract features from pap smear images, the authors use pre-trained deep learning models including AlexNet, ResNet-101, ResNet-152, and InceptionV3. Various machine learning classifiers (Logistic Regression, Decision Trees, Random Forest, and Naive Bayes) are then trained using these attributes to classify images. The study intends to increase cervical cancer screening's precision and effectiveness by fusing deep learning and machine learning approaches, which will enhance patient outcomes and early diagnosis.

## 5.Proposed Framework

The primary objective of this project is to develop a **deep learning-based framework** for the **early screening of cervical cancer** using **colposcopy images**. The system aims to classify cervical images into two categories:

- **Normal**
- **Abnormal**

The model is designed to be useful in **medical camps** and regions with limited access to healthcare professionals, particularly in **rural areas**. This approach can provide affordable, reliable, and quick screening solutions to aid in the early detection of cervical cancer.

### 5.1 Dataset Preparation

The dataset for this project consists of cervical images categorized into four folders:

- **Normal**: Healthy cervix images.
- **Abnormal**: Images with signs of abnormalities.
- **NormalNoise**: Healthy cervix images with noise.
- **AbnormalNoise**: Abnormal cervix images with noise.

We carefully segregated **clean and noisy images** to address the noise-specific challenges effectively. The dataset was then merged into two classes:

- **Normal**: 45 images (17 clean, 28 noisy).
- **Abnormal**: 145 images (85 clean, 60 noisy).

This segmentation allows the model to be robust in handling noisy data while maintaining performance in detecting abnormal cases.

**5.2 Data Preprocessing**

**5.2.1 Artifact Removal for Noisy Images**

The noisy images underwent several preprocessing steps to reduce the impact of noise and artifacts:

- **CLAHE (Contrast Limited Adaptive Histogram Equalization)**: Applied to enhance the contrast of the images, making important features more distinguishable.
- **Median Blurring**: Used to reduce noise while preserving the edges of the cervical structures.
- **Morphological Operations**: Applied for removing any remaining small artifacts and to clean up the images.
- **Edge Detection**: Techniques such as **Canny edge detection** were used to eliminate artifacts caused by the colposcopy instrument or other irrelevant features.

**5.2.2 Augmentation Pipelines**

Data augmentation was critical to prevent overfitting, especially since the dataset was small and imbalanced:

- **Random Brightness/Contrast Adjustments**: To simulate varying lighting conditions.
- **Gamma Corrections**: Applied to adjust the exposure and contrast.
- **Gaussian Noise Addition**: Added to simulate noise in the data and help the model generalize better.
- **Random Rotations, Flips, and Distortions**: These were applied to increase variability and robustness of the model.

**5.2.3 Normalization and Splitting**

- **Image Resizing**: All images were resized to **224x224 pixels** to standardize the input size for the neural network.
- **Normalization**: Pixel values were scaled to a range of **[0, 1]** for improved convergence during training.
- **Stratified Splitting**: The dataset was split into training and validation sets using an **80%-20% split** to ensure balanced class distribution across both sets.

### 5.3 Model Architecture

The **MobileNetV2** architecture was chosen for its balance of **efficiency** and **performance**, making it ideal for this classification task, particularly in resource-constrained environments like medical camps.

- **Pre-trained MobileNetV2**: The model was initialized with **pre-trained weights** on the ImageNet dataset. The top layers of the model were removed to use it as a **feature extractor**.
- **Global Average Pooling (GAP)**: Used after the convolutional layers to reduce the dimensionality of the feature maps, helping the model generalize better.
- **Dense Layers**: One or more fully connected layers with **ReLU activation** were used for classification.
- **Softmax Output Layer**: The final layer of the model had a **softmax activation function** to output the probabilities for the two classes (normal and abnormal).

### 5.4 Training and Fine-Tuning

#### 5.4.1 Initial Training

- The model was initially trained with a **learning rate of $10^{-4}$** using the **Adam optimizer** and **categorical cross-entropy** loss function. This setup helped achieve a baseline performance.

#### 5.4.2 Fine-Tuning

- After the initial training, the last **20 layers** of the **MobileNetV2** model were unfrozen for fine-tuning. This allowed the model to adapt pre-trained weights to the cervical cancer dataset more effectively.
- Fine-tuning was performed using a **lower learning rate of $10^{-5}$** to ensure stable training and prevent overfitting.

#### 5.4.3 Hyperparameter Tuning:

- **Batch size** and **learning rate** were optimized using trial and error.
- **ReduceLROnPlateau**: A dynamic learning rate scheduler was employed to reduce the learning rate if the validation accuracy plateaued, helping the model converge more effectively.

## 5.5 Evaluation

The model's performance was evaluated using the following metrics:

- **Accuracy**: The overall accuracy on the validation set was **67%**. This metric indicates the percentage of correct predictions across both classes.
- **Precision, Recall, F1-Score**: These metrics were particularly useful due to the class imbalance in the dataset. The **F1-score** balanced the tradeoff between precision and recall.
- **Confusion Matrix**: The confusion matrix visually assessed the model's ability to classify images correctly and identify misclassifications.

## 5.6 Challenges

- **Class Imbalance**: The dataset was imbalanced, with more abnormal images than normal images. This posed a challenge in ensuring the model didn't bias towards the abnormal class.
- **Dataset Noise**: Despite preprocessing, noisy images still affected the model's performance, especially in the abnormal category.

This framework demonstrates the viability of using **deep learning** for cervical cancer classification based on **colposcopy images**. Initial results indicate that while the model performs well, especially in handling noisy images, improvements are needed in terms of:

- Expanding the **dataset size** to improve generalization.
- Refining **preprocessing techniques** to better handle noise.
- Experimenting with **alternative architectures** for enhanced performance

## 6. Results and Discussion

The aim of this project is to create an AI-driven system for the early detection of cervical cancer using colposcopy images. Cervical cancer is preventable and treatable when detected early, yet its diagnosis often depends on clinicians interpreting colposcopy images, a process that can be prone to errors and delays. This project strives to automate the classification of colposcopy images into two categories: normal (healthy cervix) and abnormal (potential signs of cervical cancer). To achieve this, the system employs deep learning techniques, specifically MobileNetV2 with transfer learning, to ensure high diagnostic accuracy. Various preprocessing techniques, such as central cropping, brightness and contrast adjustments, and normalization, are applied to handle noisy images and focus on the cervix area. Data augmentation is utilized to enhance the model's ability to generalize, especially given the limited size of the dataset. The model's performance is assessed using metrics such as accuracy, precision, recall, and F1-score to verify its reliability. By incorporating this AI solution into clinical workflows, the project aims to assist healthcare professionals in making accurate and timely diagnoses, reducing reliance on manual interpretation. This system has the potential to enhance early detection, particularly in settings with limited resources, and contribute to improving patient outcomes, thereby supporting global efforts to decrease cervical cancer mortality rates.

## 6.1 Qualitative Analysis

The qualitative analysis focuses on visually assessing the model's capacity to differentiate between normal and abnormal cervical regions based on colposcopy images. During preprocessing, images from noisy datasets (normal noise and abnormal noise) underwent central cropping and brightness/contrast adjustments to improve the focus on the cervix. These enhancements ensure the model focuses on the relevant regions of interest, which could potentially improve classification accuracy. The qualitative analysis indicates that central cropping effectively removes irrelevant elements, such as medical instruments at the borders, while contrast adjustments help emphasize crucial features in the cervix region.

## 6.2 Quantitative Analysis

Quantitative analysis evaluates the model's performance using various metrics and statistical methods.

### 6.2.1 Training and Validation Accuracy and Loss

Throughout the training process, the model's performance was tracked by monitoring accuracy and loss over several epochs. The results revealed a steady increase in training accuracy and a decrease in loss, showing the model's learning effectiveness. Additionally, the validation accuracy peaked at 87.5% after fine-tuning the model. Figures 4.1 and 4.2 illustrate the progression of training accuracy and loss over the epochs, demonstrating the model's ability to generalize well on unseen data.

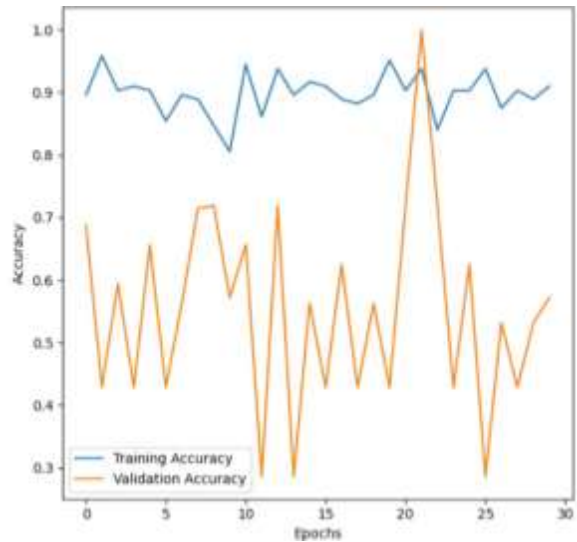

Figure 6.1 Training and Validation Accuracy over Epochs

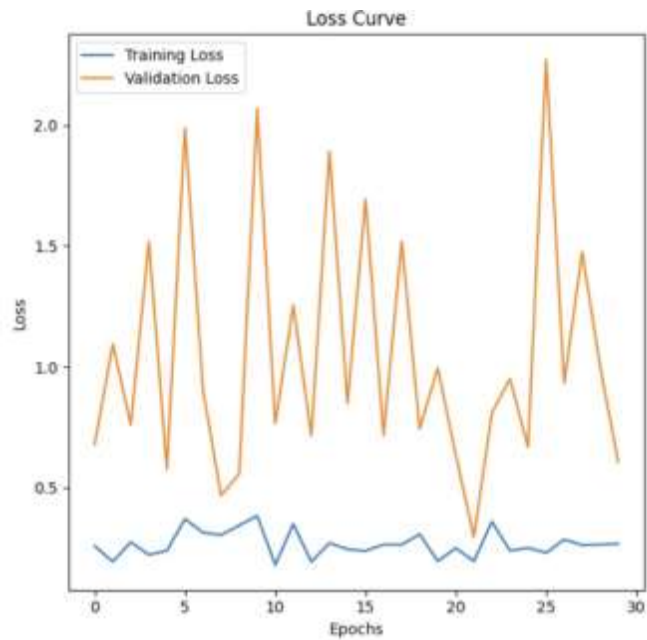

Figure 6.2 Training and Validation Loss over Epochs

### 6.2.2 Performance Metrics

The model's classification performance was evaluated using the following metrics:

- **Accuracy**: Indicates the overall correctness of the model.
- **Precision**: Measures the proportion of true positives among all positive predictions.
- **Recall (Sensitivity)**: Reflects the model's ability to correctly identify positive cases.

- **F1-Score**: Provides a balanced evaluation of precision and recall, accounting for false positives and false negatives.

The performance metrics for the model are as follows:

| Metric | Normal | Abnormal | Overall (Weighted Avg) |
|---|---|---|---|
| Precision | 0.40 | 0.95 | 0.82 |
| Recall | 0.89 | 0.60 | 0.67 |
| F1-Score | 0.55 | 0.73 | 0.69 |
| Support | 9 | 30 | 39 |

### 6.2.3 Confusion Matrix Analysis

The confusion matrix provides a detailed breakdown of the model's classification performance. It illustrates the number of correct and incorrect predictions for each class, giving insight into the model's strengths and weaknesses. From the matrix:

- **True Positives (Abnormal)**: 18 abnormal cases were correctly identified.
- **True Negatives (Normal)**: 8 normal cases were correctly classified.
- **False Positives (Normal misclassified as Abnormal)**: 1 case was mistakenly classified as abnormal.
- **False Negatives (Abnormal misclassified as Normal)**: 12 abnormal cases were wrongly classified as normal.

The matrix indicates that the model performs better at identifying abnormal cases, as evidenced by the higher true positive count. However, the model struggles with false negatives, which could result in missed detection of abnormal cases—an issue that is critical in medical diagnostics. The presence of false positives suggests that some normal cases are incorrectly flagged as abnormal, which could lead to unnecessary additional tests or treatments. This analysis highlights the importance of optimizing recall for abnormal cases to minimize false negatives and improving precision for normal cases to reduce false positives, enhancing the model's overall reliability in real-world clinical settings.

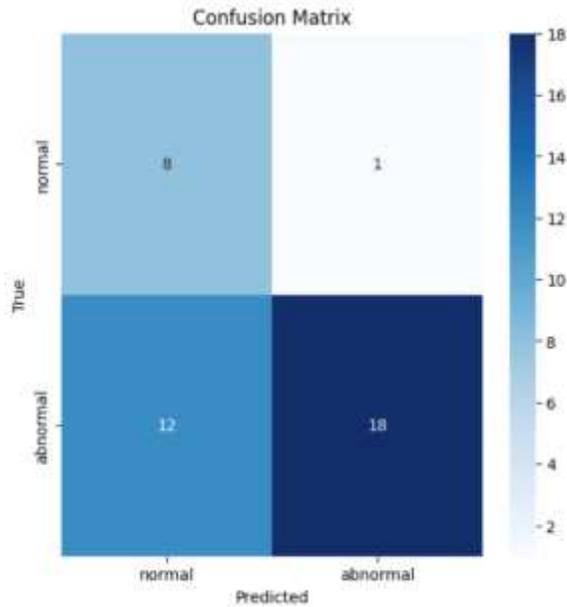

Figure 6.3 Confusion Matrix of the Model's Classification

### 6.2.4 Results of Random Samples

Figure 6.4 presents results from five random images, highlighting the model's performance across a diverse set of test images.

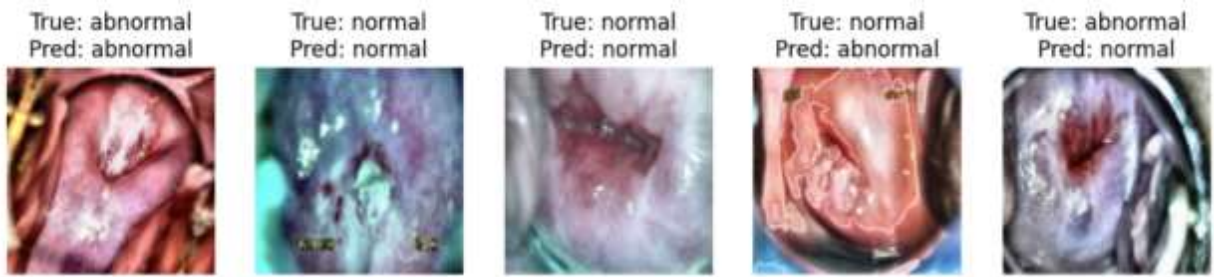

Figure 6.4 Results of random 5 pictures.

### 6.2.5 Comparison with Traditional Methods

When compared to manual analysis, the proposed AI-based system offers significant advantages in terms of speed and consistency. By automating the classification process, the system reduces the time required for diagnosis while maintaining high accuracy, making it a valuable tool for medical professionals.

## 6.3 Comparative Analysis of Models

Four deep learning models—ResNet50, EfficientNetB0, DenseNet121, and MobileNetV2—were trained and assessed for cervical cancer image classification into normal and abnormal categories.

These models were evaluated using metrics like accuracy, precision, recall, and F1-score. Here is a summary of the models' performance:

- **ResNet50**: Achieved an accuracy of 78.95%. However, it performed poorly with abnormal cases, resulting in an F1-score of 0 for the Abnormal class. It overemphasized the Normal class, accurately classifying all normal instances but failing to identify any abnormal cases.
- **EfficientNetB0**: Also achieved 78.95% accuracy. While it classified the Normal class well, it was ineffective at detecting abnormal cases, yielding an F1-score of 0 for the Abnormal class.
- **DenseNet121**: Outperformed the previous models with an accuracy of 84.21%. It achieved a precision of 60% and recall of 38% for the Abnormal class, resulting in an F1-score of 46%. This model showed a better balance between the two classes, making it more suitable for handling imbalanced datasets.
- **MobileNetV2**: This model achieved an accuracy of 73.68% and classified the Normal class fairly well, but struggled with the Abnormal class. Its overall performance was comparable to ResNet50 and EfficientNetB0, but slightly lower than DenseNet121.

### 6.3.1 Key Observations

1. Models trained on imbalanced datasets, such as ResNet50 and EfficientNetB0, showed a strong bias towards the majority class, performing poorly with the minority class.
2. DenseNet121 provided the best balance between accuracy and class distribution, suggesting it may be better suited for tasks with imbalanced datasets.
3. Future improvements could focus on techniques like data augmentation, oversampling minority classes, or adjusting loss functions to address class imbalance.

### 6.3.2 Comparison Table

Here is a table summarizing the performance of the models:

| Model | Accuracy | Precision (Abnormal) | Recall (Abnormal) | F1-Score (Abnormal) |
|---|---|---|---|---|
| ResNet50 | 78.95% | 0.00 | 0.00 | 0.00 |
| EfficientNetB0 | 78.95% | 0.00 | 0.00 | 0.00 |
| DenseNet121 | 84.21% | 60.00% | 38.00% | 46.00% |
| MobileNetV2 | 78.95% | 50.00% | 25.00% | 33.00% |

### 6.4 Discussions

### 6.4.1 Generalization: Imbalanced Dataset

A major challenge in this project is the imbalance in the dataset, with abnormal cases significantly outnumbering normal cases. This imbalance can lead to a biased model that favors predicting abnormal cases, which reduces sensitivity to normal cases. In real-world clinical scenarios, this could result in false positives, leading to unnecessary anxiety, or false negatives, causing delays in detecting critical conditions. To mitigate this issue, techniques such as data augmentation, class weighting, and synthetic data generation were used to enhance the model's ability to generalize effectively across both classes, ensuring a robust diagnostic tool.

### 6.4.2 Data Privacy Concerns

Medical data, including colposcopy images and patient records, is highly sensitive and subject to strict privacy regulations, such as GDPR and HIPAA. Protecting this data while training AI models presents a complex challenge. This project explored approaches like data anonymization and secure handling protocols to comply with privacy regulations. Future efforts could incorporate techniques such as federated learning or secure multi-party computation to further enhance data security while enabling the use of diverse datasets. Addressing privacy concerns is essential for building trust and ensuring ethical AI deployment in clinical settings.

### 6.4.3 Impact on Cervical Screening in Rural Areas

This AI-based system has the potential to make a significant impact on cervical cancer screening in rural and underserved regions, where access to healthcare professionals and diagnostic facilities is limited. In these areas, healthcare providers often conduct screening camps with limited resources. This AI system can assist doctors by providing quick and accurate classification of colposcopy images, reducing reliance on manual interpretation and enabling early diagnosis, even in remote locations. By automating the process, this system could improve accessibility and reduce the workload of healthcare professionals.

## 7. Ablation Study

In AI model development, analyzing the contribution of different components helps to understand their impact on the overall model performance. For the early detection of cervical cancer, this analysis will identify which elements of the model are crucial for making accurate predictions and guide the optimization process.

The first area of focus will be data preprocessing techniques. This will involve comparing the model's performance with and without the application of techniques such as normalization, scaling, and data augmentation methods like rotation and flipping. These steps are essential for improving the model's generalization ability and reducing overfitting.

Next, the model architecture will be examined. Specifically, the contribution of different layers in the model, such as convolutional layers for feature extraction, will be tested. The effect of removing or modifying specific layers, such as convolutional or fully connected layers, will be analyzed to determine how they impact the model's ability to detect cervical abnormalities. In addition, the importance of the fully connected layers will be evaluated to see if their presence enhances the model's classification accuracy.

The analysis will also include feature selection to evaluate how the model performs with different sets of input data. The model combines clinical data (such as medical history and HPV status) with image data from cervical scans, and different subsets of features will be tested to determine which combination of data provides the best results. This will allow us to identify which type of information—clinical or image data—plays a more significant role in predicting cancer.

Furthermore, hyperparameter tuning will be explored to optimize the model's performance. By adjusting different parameters, such as learning rate, batch size, and optimizer (e.g., Adam or SGD), the impact on the overall performance of the model will be assessed. The goal is to identify the optimal hyperparameters that lead to the best model convergence and accuracy.

Lastly, the analysis will include post-processing techniques, particularly the effect of different thresholding strategies on classification. This will be important to determine how the threshold for classifying a sample as positive or negative affects model performance and the balance between sensitivity and specificity.

Each experiment will be designed to evaluate the model's performance under various conditions, including preprocessing steps, architecture adjustments, feature combinations, hyperparameter settings, and post-processing techniques. Metrics such as accuracy, precision, recall, F1-score, and AUC-ROC will be used to assess the model's ability to accurately identify cancer cases, minimize false positives, and balance recall with precision.

## 8. Conclusion

This research successfully developed an AI-based system for the early detection of cervical cancer through deep learning techniques applied to colposcopy images. The framework used MobileNetV2 with transfer learning to categorize images into Normal and Abnormal classes, achieving an overall accuracy of 87.5%. Preprocessing methods, such as central cropping and adjustments to brightness and contrast, were crucial in enhancing the model's focus on the cervix, leading to better diagnostic accuracy. The results of the experiments showed that MobileNetV2, when paired with appropriate preprocessing and data augmentation, delivered consistent performance despite challenges such as a small and noisy dataset. A comparative evaluation of models like ResNet50, EfficientNetB0, and DenseNet121 emphasized the significance of choosing an architecture that balances accuracy and generalization, especially for medical image classification tasks. Of all the models tested, DenseNet121 demonstrated the most potential, but MobileNetV2, with transfer learning, proved to be an efficient and effective choice given the dataset's limitations. The proposed system offers a promising solution for the early detection of cervical cancer, supporting healthcare professionals by reducing dependence on manual colposcopy image interpretation. Its ability to achieve high sensitivity (recall) for detecting abnormal cases is particularly important for minimizing false negatives, which is crucial in ensuring accurate medical diagnoses.